\documentclass[final,3p,sort&compress]{elsarticle}

\usepackage{graphicx}

\usepackage{amsmath,amssymb,amsfonts}
\usepackage{bm}

\def\be{\begin{equation}}
\def\ee{\end{equation}}
\def\ba{\begin{eqnarray}}
\def\ea{\end{eqnarray}}
\def\bd{\begin{displaymath}}
\def\ed{\end{displaymath}}
\def\bq{\begin{eqnarray}}
\def\eq{\end{eqnarray}}

\journal{Annals of Physics}

\begin{document}

\begin{frontmatter}



\title{Dissipative Bohmian mechanics within the Caldirola-Kanai framework:\\
A trajectory analysis of wave-packet dynamics in viscid media}


\author[label1]{A. S. Sanz\corref{corresp}} \ead{asanz@iff.csic.es}
\author[label2]{R. Mart\'{\i}nez-Casado\fnref{newaddress}}
\author[label3]{H. C. Pe\~{n}ate-Rodr\'{\i}guez}
\author[label3]{G. Rojas-Lorenzo}
\author[label1]{S. Miret-Art\'es}

\cortext[corresp]{Corresponding author}
\fntext[newaddress]{Present address: Departamento de F{\'\i}sica
Te{\'o}rica de la Materia Condensada, Universidad Aut{\'o}noma de
Madrid, 28049 Madrid, Spain}

\address[label1]{Instituto de F\'{\i}sica Fundamental (IFF-CSIC),
Serrano 123, 28006 Madrid, Spain}

\address[label2]{Department of Chemistry, Imperial College London,
South Kensington, London SW7 2AZ, United Kingdom}

\address[label3]{Instituto Superior de Tecnolog\'{\i}as y Ciencias
Aplicadas, Ave.\ Salvador Allende y Luaces, Quinta de Los Molinos,
Plaza, La Habana 10600, Cuba}

\begin{abstract}
Classical viscid media are quite common in our everyday life.
However, we are not used to find such media in quantum mechanics, and
much less to analyze their effects on the dynamics of quantum systems.
In this regard, the Caldirola-Kanai time-dependent Hamiltonian
constitutes an appealing model, accounting for friction without
including environmental fluctuations (as it happens, for example,
with quantum Brownian motion).
Here, a Bohmian analysis of the associated friction dynamics is
provided in order to understand how a hypothetical, purely quantum
viscid medium would act on a wave packet from a (quantum) hydrodynamic
viewpoint.
To this purpose, a series of paradigmatic contexts have been chosen,
such as the free particle, the motion under the action of a linear
potential, the harmonic oscillator, or the superposition of two
coherent wave packets.
Apart from their analyticity, these examples illustrate interesting
emerging behaviors, such as localization by ``quantum freezing'' or
a particular type of quantum-classical correspondence.
The reliability of the results analytically determined has been
checked by means of numerical simulations, which has served to
investigate other problems lacking of such analyticity (e.g., the
coherent superpositions).
\end{abstract}



\end{frontmatter}



\section{Introduction}
\label{sec1}

An iron ball falling in oil reaches after some time a constant
velocity.
It is then not accelerated anymore. The same
happens with rain drops in air.
These are two examples of our everyday
life where a classical system undergoes a uniform motion when
acted by a viscid medium (oil in the first example and air in the
second one).
This effect arises when the friction with such a medium
compensates the acceleration induced by the gravity acting on the
system. But, what about quantum systems? May they display such
behaviors? Actually, how would a viscid quantum medium act on a
quantum system, if such a medium exists? These are natural
questions that come to our minds when trying to establish a
correspondence with the analogous classical systems.
However, before considering a model to describe such a situation,
it is interesting to make some considerations on quantum systems
in the light of the theory of open quantum systems
\cite{weiss-bk,breuer-bk:2002}.

Real systems are not in isolation in Nature. Rather, we find them
coupled to other systems, namely an
environment or bath. The usual way to tackle their study is first by
considering a system-plus-reservoir Hamiltonian system, which
includes all the involved degrees of freedom (those from the system
plus those from the environment) as well as their coupling. However,
not always a full quantum-mechanical description in these terms is
affordable and therefore one has to consider simpler models consisting
of the bare (isolated) system Hamiltonian plus an effective,
time-dependent interaction.
Although these models do not provide us with any information regarding
the environmental dynamics, they are very convenient to describe its
effects on the system, even if the problem becomes
{\it nonconservative}.
Phenomenological equations only
accounting for the system dynamics are, for example, the Lindblad or
the quantum Langevin equations. In the first case, an effective
description of the evolution of the (reduced) system density matrix
is achieved by including a series of dissipation operators or
dissipators in the corresponding equation of motion. In the second
case, the equation describes the evolution of an operator associated
with a certain observable (e.g., the system position), and apart
from a dissipative term the equation also includes a stochastic
noise related to the environmental fluctuations. Nevertheless, in
both cases the effect on the system dynamics is the same: a loss of
the system coherence (decoherence) and a relaxation or damping of
its energy.

In the particular case of the quantum Langevin equation, a (quantum)
noise function is included in order to account for the environmental
thermal fluctuations on the system (Brownian-like motion).
This equation can be reformulated in terms of a Hamiltonian model,
where the noise arises from a collection of harmonic oscillators
coupled to the system.
This is the so-called Caldeira-Leggett model
\cite{caldeira:PhysicaA:1983,caldeira:AnnPhys:1983}.
Predating this model, though, we find a former dissipative
one, the so-called Caldirola-Kanai model
\cite{caldirola:NuovoCim:1941,kanai:ProgTheorPhys:1948}.
This is a Hamiltonian reformulation of the Langevin equation with zero
fluctuations, i.e., the Brownian-like thermal fluctuations that sustain
the system motion are neglected and the system undergoes a gradual
decay until all its energy is completely, irreversibly lost by
dissipation.
Then the system dynamics stops.

The behavior exhibited by quantum systems described by the
Caldirola-Kanai model contradicts the uncertainty relations ---even
so it has been considered by a number of authors with different
purposes, from computational issues to more fundamental aspects of
dissipative quantum dynamics
\cite{kerner:CanJPhys:1958,schuch:IJQC:1999,majima:AnnPhys:2011,%
jannussis:PhysicaA:1980,jannussis:PhysLett:1979,khandekar:JMathPhys:1979,%
lorenzen:PRA:2009,sun:PRA:1995,sun:PRA:1994,schuch:PRA:1997,schuch:JPA:2001,%
schuch:RevMexFis:2001,bowman:JCP:1981,guerra:PhysRep:1981}.
The reason for this behavior comes from the fact that this model is
incomplete, i.e., it is a model that only describes a system gradually
losing its energy, but not a {\it larger} model that includes another
subsystem gaining the energy that our system loses
\cite{bateman:PhysRev:1931,brittin:PhysRev:1950,senitzky:PhysRev:1960,greenberger:JMathPhys:1979,sanz-bk-1}.
To avoid this inconvenience, alternative nonlinear Schr\"odinger
equations have been proposed in the literature, one of them being
Kostin's model \cite{kostin:jcp:1972}, which is somehow connected with
the Bohmian approach that is considered in this work.
Contrary to the Caldirola-Kanai model, Kostin's Hamiltonian does not
explicitly depend on time, although it contains a friction term
proportional to the phase of the wave function (and hence an implicit
time-dependence).
Within the Bohmian or hydrodynamic picture of quantum mechanics
\cite{madelung:ZPhys:1926}, this term is thus related to the local
value of the velocity of an element of the quantum fluid (i.e., the
quantum system treated as a fluid), thus establishing a connection
between the classical concept of friction (as it appears in the
Langevin equation) and a feasible quantum one.

The purpose of this work is to analyze the dissipative dynamics
associated with the Caldirola-Kanai model in the framework of Bohmian
mechanics (quantum hydrodynamics) to shed some light on the dynamics
affected by a hypothetical, fully quantum viscid medium.
As pointed out above, the system dynamics will exhibit full
dissipation (which can be considered a pathology of the model, but
an interesting one), because its incompleteness, i.e., the lack of
fluctuations that allow the system to reach some equilibrium
\cite{cavalcanti:PRE:1998}, as happens in quantum Brownian motion,
or another subsystem gaining the energy that it loses
\cite{bateman:PhysRev:1931}.
In this regard, the trajectories or streamlines obtained should be
considered as reduced ones \cite{sanz:EPJD:2007,sanz:CP:2011}, i.e.,
solely related to the system studied by with no relation to any larger
system.
In this sense, these trajectories become tools to only describe the
(reduced) dynamics of the subsystem under consideration.
By inspecting the topology displayed by the Bohmian trajectories, the
effects of dissipation readily become apparent, such as localization
by freezing the wave function evolution or the emergence of
classical-like behaviors.

As far as we know, Bohmian mechanics has not been widely used within
the context considered here despite of the fact that it seems to be
an appropriate tool to analyze the quantum dissipative phenomena. In
this regard, the first application was carried out about 20 years
ago by Vandyck \cite{vandyck:JPA:1994}, who tackled the issue of the
decay of harmonic oscillator eigenstates. Ten years later, Shojai
and Shojai analyzed \cite{shojai:PRAMANA:2004} the problem of
friction of two-level system transition by reformulating such a
problem in Bohmian terms. At a more practical level, Tilbi {\it et
al.}\ \cite{tilbi:PhysScr:2007} used Bohmian mechanics to derive an
expression of the Caldirola-Kanai propagator starting from the
Feynman path integral approach.
In this regard, apart from offering a Bohmian description of
dissipative systems within the Caldirola-Kanai context, another purpose
of this work is to establish a general hydrodynamic dissipative
framework (within this model) where the use of transformations that
make the system to satisfy the uncertainty relations
\cite{schuch:IJQC:1999,schuch:PRA:1997,schuch:JPA:2001,schuch:RevMexFis:2001}
is not necessary.
This is essentially in view of tackling more complex, higher-dimensional
systems, where such transformations cannot be easily found (if possible
at all) and the use of numerical simulations is mandatory.
These further analyses, though, have been left out of
the scope of the present work, since the aim here is to investigate
analytically (as far as possible) the friction effects on the
quantum system by means of the corresponding Bohmian trajectories.
In this sense, we have considered simple yet physically insightful
scenarios, such as the wave-packet dynamics in free space, and also
under the action of linear and harmonic potentials, as well as two
wave-packet interference, all of them affected by the friction of a
surrounding viscid medium. These examples will allow us to
understand how friction gives rise to localization, to a type of
quantum-to-classical transitions, or to the disappearance of the
zero-point energy.

Taking into account these scopes, the work has been organized as
follows. A brief overview of the
Caldirola-Kanai model, its quantization, and the formulation of the
associated Bohmian mechanics is presented in Sec.~\ref{sec2}. In
Sec.~\ref{sec3} we introduce the elements to proceed our analytical
study, in particular, the use of a Gaussian ansatz which depends on
the canonical position and momentum, and whose evolution can be
determined by solving a set of ordinary differential equations, thus
simplifying the Bohmian analysis. The results and discussion arising
from the application of this approach to the scenarios mentioned
above are accounted for in Sec.~\ref{sec4}.
More specifically, the cases of a free motion, motion under a linear
potential, and the harmonic potential have been treated.
The analytical results obtained by means of the method described in
Sec.~\ref{sec3} are in agreement with those formerly found by Hasse
\cite{hasse:JMathPhys:1975}, as well as with numerical simulations
performed to check their reliability and feasibility for further
extension to more complex situations.
Finally, in Sec.~\ref{sec5} a series of concluding remarks drawn from
this work are discussed.


\section{Dissipative Bohmian mechanics}
\label{sec2}

There are different strategies to construct Hamiltonian functions
to account for dissipative systems and then, from them, to obtain
the corresponding quantum Hamiltonian operators
\cite{razavy-bk,kochan:pra:2010,sanz-bk-1}.
One of them consists in considering
time-dependent Lagrangians (and therefore Hamiltonians), which avoids
us to deal explicitly with the environmental degrees of freedom.
This approach preserves the canonical formalism and can be a good
starting point to find
out the quantal analog of the corresponding dissipative
dynamics. Here, the Caldirola-Kanai Hamiltonian model constitutes
a paradigm of dissipation.
This model arises
from the classical equation of motion for a damped particle of
mass $m$ under the action of an external potential $V(x)$ (for
simplicity and practical purposes, we will consider
one-dimensional models throughout this work) and a mean friction
$\gamma$,
\be
 m\ddot{x} + m\gamma\dot{x} + \partial_x V(x) = 0 ,
 \label{langed}
\ee
with the short-hand notation $\dot{x}=dx/dt$ and
$\ddot{x}=d^2x/dt^2$ for total time derivatives, and
$\partial_\eta^k = \partial^k/\partial \eta^k$ for the
$k$th-derivative with respect to a given variable $\eta$.
Multiplying this equation by $e^{\gamma t}$, it can be recast as
\be
 \frac{d}{dt}\left( m e^{\gamma t} \dot{x} \right)
  + \partial_x \left[ V(x) e^{\gamma t} \right] = 0 .
 \label{lagr-eq}
\ee
If we consider the change of variable $X=x$ and
\be
 P \equiv m e^{\gamma t} \dot{x} = p e^{\gamma t} ,
 \label{momentumd}
\ee
where $p = m\dot{x}$ is the physical momentum, we readily notice that
(\ref{lagr-eq}) is just the Lagrange equation satisfied by the
time-dependent Lagrangian function
\be
 \mathcal{L} =\left[ \frac{1}{2}\ m\dot{X}^2 - V(X) \right] e^{\gamma t}
  = \frac{P^2}{2m}\ e^{-\gamma t} - V(X) e^{\gamma t} ,
 \label{langd}
\ee
where $P$ plays the role of a canonical momentum, i.e.,
$P = \partial_{\dot{X}} \mathcal{L}$.
This equation allows us to obtain straightforwardly the corresponding
Hamiltonian,
\be
 \mathcal{H} = \dot{X} P - \mathcal{L}
  = \frac{P^2}{2m}\ e^{-\gamma t} + V(X) e^{\gamma t} ,
 \label{hamCK}
\ee
which is a function of the canonical variables $X$ and $P$, satisfying
the Hamilton equations
\be
 \dot{X} = \partial_P \mathcal{H} , \qquad
 \dot{P} = -\partial_X \mathcal{H} ,
 \label{hamflow}
\ee
respectively.
Equation~(\ref{hamCK}) defines the Caldirola-Kanai
Hamiltonian\footnote{As it was shown by Yu and Sun
\cite{sun:PRA:1995,sun:PRA:1994}, this model can be obtained by
considering a collection of harmonic oscillators coupled to the
system with only dissipation (no noise) and described by the
coordinates $X$ and $P$. In terms of $x,p$, the corresponding
equation of motion for the coordinate is right, but the same is not
true for $p$}, denoted by $\mathcal{H}$.
In terms of this time-dependent Hamiltonian model, the system classical
energy can be recast as
\be
 E = \frac{p^2}{2m} + V(x) = \mathcal{H} e^{-\gamma t} ,
 \label{energyd}
\ee
which explicitly exhibits the typical exponential decay associated with
Eq.~(\ref{langed}) ---regardless of the implicit time-dependence that
$\mathcal{H}$ might display.

The quantum analog of (\ref{hamCK}) can be now obtained by considering
the (canonical) momentum operator $\hat{P}=-i\hbar\partial/\partial X$,
which leads to the standard commutation rule, $[\hat{X},\hat{P}] =
i\hbar$ ---to appreciate the difference between the canonical and
physical variables, compare this relation to the one for the latter
variables, $[\hat{x},\hat{p}] = i\hbar e^{-\gamma t}$, which implies
the violation of the uncertainty principle.
The quantum Hamiltonian operator thus reads as
\be
 \hat{\mathcal{H}} = - \frac{\hbar^2}{2m}\ e^{-\gamma t}\
  \partial^2_X + V(X) e^{\gamma t} .
 \label{hamCKQ}
\ee
Now, given the dependence of this Hamiltonian {\it only} on $X$,
it can be recast in terms of $x$, without loss of generality.
This leads us to the the generalized, dissipative Schr\"odinger
equation,
\be
 i\hbar\partial_t \Psi = -\frac{\hbar^2}{2m}\ e^{-\gamma t}
  \partial^2_x \Psi + V(x) e^{\gamma t} \Psi ,
 \label{dissschro}
\ee
in terms of the physical coordinate.
Notice that as long as quantities related to the momentum operator are
not computed, the use of this wave equation in the physical coordinate
space is formally correct (otherwise, the canonical operators $\hat{P}$
and $\hat{X}$ should be considered).
Hence, at this stage, we can consider the polar ansatz
$\Psi(x,t) = \rho^{1/2}(x,t) e^{iS(x,t)/\hbar}$, typically used in
Bohmian mechanics, where $\rho$ is the probability density and $S$ is
the real phase field.
If this ansatz is substituted into (\ref{dissschro}), we obtain
\ba
 \partial_t \rho & + & \partial_x \mathcal{J} = 0 , \\
 \partial_t S & + & \frac{(\partial_x S)^2}{2m}\ e^{-\gamma t}
   + \mathcal{V}(x,t)_{\rm eff} = 0 ,
\ea
where $\mathcal{J} = \rho (\partial_x S/m) e^{-\gamma t}$ is the
associated probability density current and $\mathcal{V}_{\rm eff}(x,t)
= V(x) e^{\gamma t} + \mathcal{Q}(x,t)$ is an effective potential,
which includes the quantum potential
\be
 \mathcal{Q} = - \frac{\hbar^2}{2m}
   \frac{\partial_x^2 \rho^{1/2}}{\rho^{1/2}}\ e^{-\gamma t}
   = - \frac{\hbar^2}{4m} \left[
    \frac{\partial_x^2 \rho}{\rho} - \frac{1}{2}
   \left( \frac{\partial_x \rho}{\rho} \right)^2 \right]
    e^{-\gamma t} .
\ee
As it can be noticed, both $\mathcal{J}$ and $\mathcal{Q}$ display
the same functional form as their non-dissipative counterparts, but
multiplied by the time-dependent decaying factor, $e^{-\gamma t}$.
This fact also manifests in the corresponding dissipative Bohmian
trajectories, which are obtained from the equation of motion
\be
 \dot{x} = \frac{\mathcal{J}}{\rho}
   = \frac{\partial_x S}{m}\ e^{-\gamma t}
   = \frac{\hbar}{2mi\rho}
   \left( \Psi^* \partial_x \Psi - \Psi \partial_x \Psi^* \right)
     e^{-\gamma t} .
 \label{bohmtrad}
\ee
Note how this expression resembles Eq.~(\ref{momentumd}), although it
has been directly obtained in terms of the physical coordinates.
That is, the correct expression for the Bohmian momentum is
$P_B = \partial_X S = m e^{\gamma t} \dot{X}$, from which the Bohmian
trajectories in the physical space are obtained by substituting $X$ by
$x$, which leads to the expression on the right-hand side of the
second equality in Eq.~(\ref{bohmtrad}).


\section{Analytical wave-packet propagation}
\label{sec3}

In principle, general quantum dissipative problems described by
the Hamiltonian model (\ref{hamCKQ}) have to be tackled by means
of numerical techniques, where the dissipative, time-dependent
Schr\"odinger equation (\ref{dissschro}) can be solved by using
standard wave-packet propagation methods. Nevertheless, as also
happens with non-dissipative quantum mechanics, for potential
functions which are of quadratic or lesser order, we can find
analytic solutions, which turn out to be very insightful to better
understand the physics of some problems. In particular, this is
the case for wave functions that are initially described by
Gaussian wave packets. These solutions can be easily obtained by
proceeding as follows. It is well-known that a few parameters are
enough to completely characterize a wave function described by a
Gaussian wave packet \cite{heller:JCP:1975,tannor-bk}. So, for
analytical purposes, consider the ansatz \cite{sanz-bk-2}
\be
 \Psi(X,t) = e^{(i/\hbar)\left[ \alpha_t (X-X_t)^2
   + P_t(X-X_t) + f_t\right]} ,
 \label{waved}
\ee
where a subscript $t$ is used throughout this work to denote
explicit time-dependence of the parameters characterizing the wave
function (the subscript 0 indicates their value at $t=0$).
This ansatz has been expressed in terms of the canonical
variables, because it involves the momentum and, as mentioned above,
only the canonical momentum makes meaningful the Schr\"odinger
equation (\ref{dissschro}) ---the kinetic operator was defined
according to $\hat{P}$, not with respect to the physical momentum, $p$.
Nevertheless, later on it will be recast in terms
of the physical variables, which represent the correct (physical)
centroidal trajectory ($x_t,p_t$), in order to analyze the friction
effects in the real configuration space.
Regarding the shape and normalization complex functions $\alpha_t$
and $f_t$, since they are eventually functions of the canonical
coordinates, they can be readily recast in terms of the physical ones.
Therefore, it is not necessary to make any additional assumption on
them.

Consider that the wave function (\ref{waved}) is normalized, which
means that the imaginary part of $f_t$ normalizes this wave function
and follows from assuming the initial condition
\be
 f_0 = \frac{i\hbar}{4}\
  \ln \left[ \frac{\pi\hbar}{2{\rm Im}\{\alpha_0\}} \right] ,
\ee
with ${\rm Im}\{\alpha_0\} \neq 0$. In such a case, the position
and momentum expectation values (and therefore the wave-packet
centroid) follow a classical trajectory, i.e., $\langle \hat{X}
\rangle(t) = X_t$ and $\langle \hat{P} \rangle(t) = P_t$, with
($X_t,P_t$) being obtained by integrating the Hamilton equations
of motion (\ref{hamflow}), i.e.,
\ba
 \dot{X}_t & = & \frac{P_t}{m}\ e^{-\gamma t} ,
 \label{wave1} \\
 \dot{P}_t & = & - \partial_{X_t} V_t e^{\gamma t} ,
 \label{wave2}
\ea
with $V_t \equiv V(X_t)$.
Thus, the physical dispersion of the wave packet is found to be
\be
 \Delta x(t) = \Delta X(t) =
  \sqrt{\langle X^2 \rangle(t) - [\langle X \rangle(t)]^2}
  = \sqrt{\frac{\hbar}{4{\rm Im}\{\alpha_t\}}} = \sigma_t ,
\ee
where $\sigma_t$ is the instantaneous wave-packet spreading (see
Sec.~\ref{sec4-1}).
The validity of this expression arises from the fact that the
transformation from $X$ to $x$ is general and does not explicitly
depend on the particular time-dependence displayed by $X$.
Regarding the energy expectation value, it is given by
\ba
 \bar{E} = \langle \hat{\mathcal{H}} \rangle e^{-\gamma t} & = &
  \frac{P_t^2}{2m}\ e^{-2\gamma t} + V_t
  + \frac{\hbar}{2m} \frac{|\alpha_t|^2}{{\rm Im}\{\alpha_t\}}\
    e^{-2\gamma t}
  + \frac{\hbar V''_t}{8{\rm Im}\{\alpha_t\}}
  \nonumber \\
 & = & E_t
  + \frac{\hbar}{2m} \frac{|\alpha_t|^2}{{\rm Im}\{\alpha_t\}}\
    e^{-2\gamma t}
  + \frac{\hbar V''_t}{8{\rm Im}\{\alpha_t\}} ,
\ea
where $E_t$ is the classical energy evaluated along the trajectory
$(x_t,p_t)$ solution of (\ref{langed}) (see Eq.~(\ref{energyd})).
In this sense, $\bar{E}$
can be split up into two contributions, one coming from the
translational motion along the classical centroidal trajectory
($E_t$) and another one related to the wave-packet spreading
\cite{sanz:JPA:2008}. More specifically, this latter contribution
contains information about the spatial variations of both the wave
packet and the potential.

Regarding the shape parameters $\alpha_t$ and $f_t$, their
equations of motion can be readily obtained as follows. Let us
first recast $V(X)$ as a Taylor power series around the centroidal
position $X_t$ up to the second order, i.e.,
\be
 V(X) = V_t + V'_t (X - X_t) + \frac{1}{2}\ V_t''(X-X_t)^2
 \label{vexp}
\ee
where the primes denote the order of the derivative with respect
to $X$ and the subscript $t$ the evaluation at $X_t$ (and the
implicit dependence on time). Substituting the ansatz
(\ref{waved}) and the expansion (\ref{vexp}) into the dissipative
Schr\"odinger equation (\ref{dissschro}), and then comparing the
coefficients associated with the same power of $(X-X_t)$, it is
found that
\ba
 \dot{\alpha}_t & = & - \frac{2\alpha_t^2}{m}\ e^{-\gamma t}
  - \frac{1}{2}\ V''_t e^{\gamma t} ,
 \label{wave3}
 \\
 \dot{f}_t & = &
  \frac{i\hbar\alpha_t}{m}\ e^{-\gamma t} + \mathcal{L}_t ,
 \label{wave4}
\ea
with $\mathcal{L}_t$ given by (\ref{langd}) evaluated along the
classical trajectory $(X_t,P_t)$ ---since this term is only
time-dependent and does not have any space dependence (see
Sec.~\ref{sec4-1}), it will not influence the Bohmian dynamics and
therefore we will not consider its explicit functional form. Thus,
integrating the set of coupled ordinary differential equations
(\ref{wave1}), (\ref{wave2}), (\ref{wave3}) and (\ref{wave4}), we
have the wave function (\ref{waved}) completely specified at any
time. Substituting this wave function into the Bohmian equation of
motion (\ref{bohmtrad}), the following expression for the velocity
is readily reached
\be
 \dot{X} = \left[ \frac{P_t}{m}
  + \frac{2{\rm Re}\{\alpha_t\}}{m}\ (X-X_t) \right] e^{-\gamma t} ,
 \label{bohmdH}
\ee
which after integration renders the corresponding dissipative
Bohmian trajectory $x(t) = X(t)$. Equation~(\ref{bohmdH}) turns
out to be  very interesting, because if the second term vanishes,
we find that the Bohmian trajectory exactly coincides with the
classical one given by (\ref{wave2}). Notice that, accordingly,
the condition of classicality does not require necessarily of the
limiting procedure of  $\hbar\to 0$, but that the wave packet
remains relatively localized (i.e., $1/{\rm Re}\{\alpha_t\} \to
0$), in agreement with the hypothesis of Ehrenfest's theorem
\cite{sanz:AJP:2012}. On the other hand, this equation also shows
us that quantum motion can be observed even though the probability
density becomes negligible provided the phase is well-defined. In
other words, if ${\rm Im}\{\alpha_t\} \to 0$ and the wave function
becomes a pure phase factor (except for some time-dependent norm
coming from $f_t$), quantum motion can still be observed through
both $P_t$ and ${\rm Re}\{\alpha_t\}$. Later on, some example of
such a behavior is discussed.

In the examples studied in the next Section, we are going to consider
as an initial wave function the Gaussian wave packet
\be
 \Psi(x,0) = \left(\frac{1}{2\pi\sigma_0^2}\right)^{1/4}
   e^{-(x-x_0)^2/4\sigma_0^2 + ip_0(x-x_0)/\hbar} .
 \label{wave0}
\ee
This means that the initial conditions for the above parameters and
variables will be $X_0=x_0$, $P_0=p_0$, $\alpha_0=i\hbar/4\sigma_0^2$,
and $f_0=(i\hbar/4)\ln(2\pi\sigma_0^2)$.
Regarding the Bohmian trajectories, we have considered sets of
initial positions distributed according to the initial probability
distribution,
\be
 \rho(x,0) = \frac{1}{\sqrt{2\pi\sigma_0^2}}\
   e^{-(x-x_0)^2/2\sigma_0^2} ,
 \label{rho0}
\ee
in order to better visualize the friction effects on different regions
of the wave packet.
In order to test the reliability of our analytical derivations following
the procedure based on the Gaussian ansatz (\ref{waved}), the results
shown arise from numerical simulations.
As it can be seen, concerning the wave-packet evolution, both
analytical derivations and numerical results are in agreement with
previous results obtained for the same Caldirola-Kanai model by Hasse
\cite{hasse:JMathPhys:1975}.


\section{Applications}
\label{sec4}


\subsection{Free propagation}
\label{sec4-1}

Consider a free Gaussian wave packet. After integration of the
aforementioned equations of motion, we have that
\ba
 x_t & = & x_0 + \frac{p_0}{m\gamma}\ (1 - e^{-\gamma t}) ,
 \label{free1} \\
 p_t & = & p_0 e^{-\gamma t} ,
 \label{free2} \\
 \alpha_t & = & \frac{\alpha_0}{1 + (2\alpha_0/m\gamma)
  (1- e^{-\gamma t})}
 \label{free3} \\
 f_t & = & \frac{i\hbar}{4}\
  \ln \left[ \frac{\pi\hbar}{2{\rm Im}\{\alpha_0\}} \right]
  + \frac{i\hbar}{2}
  \ln\left[1 +
    \frac{2\alpha_0}{m}\left(\frac{1 - e^{-\gamma t}}{\gamma}\right)\right]
  + \mathcal{S}_{{\rm cl},t} ,
 \label{free4}
\ea
where
\be
 \mathcal{S}_{{\rm cl},t} = \int_0^t \mathcal{L}_{t'} dt'
\ee
is the associated classical action. As mentioned in the previous
section, this term only adds a time-dependent phase factor, which
does not play any role in the Bohmian dynamics (its gradient
vanishes). Equation~(\ref{free3}) can be alternatively expressed
in terms of an effective time-dependent spreading
\cite{sanz:JPA:2008}
\be
 \tilde{\sigma}_t = \sigma_0 \left[ 1 + \frac{i\hbar}{2m\sigma_0^2}
     \left(\frac{1-e^{-\gamma t}}{\gamma}\right) \right] ,
 \label{freespread}
\ee
which is connected to $\alpha_t$ by the simple relationship
$\alpha_t = i\hbar/4\sigma_0\tilde{\sigma}_t$.
Accordingly, Eq.~(\ref{free4}) can be recast in a simpler form,
\be
 f_t = \frac{i\hbar}{4}\ \ln (2\pi\tilde{\sigma}_t^2)
  + \mathcal{S}_{{\rm cl},t} .
\ee
It is worth mentioning that, in terms of the canonical variables, the
functional form of this wave packet is the same as that displayed by
a standard free wave packet if $t$ in the latter is replaced by
$\tau = (1 - e^{-\gamma t})/\gamma$ in the former
\cite{schuch:RevMexFis:2001}, i.e., a time contraction coming from
the dissipative process and associated with the eventual ``freezing''
(see below) displayed by the wave packet.

In the limit $\gamma\to 0$, it is straightforward to show that
(\ref{waved}) approaches the well-known solution for the free Gaussian
wave packet \cite{sanz:JPA:2008},
\begin{equation}
 \Psi (x,t) = \left[\frac{1}{2\pi(\tilde{\sigma}_t^{0})^2}\right]^{1/4}
  e^{-(x-x_t)^2/4\sigma_0\tilde{\sigma}_t^0
   + ip_0(x-x_t)/\hbar + iEt/\hbar} ,
 \label{wfsanz08}
\end{equation}
where $x_t = x_0 + (p_0/m)t$ denotes the instantaneous centroidal
position, $E = p_0^2/2m$ is its total mechanical energy, and
\be
 \sigma_t^0 = |\tilde{\sigma}_t^0|
   = \sigma_0\sqrt{1 + \left( \frac{\hbar t}{2m\sigma_0^2} \right)^2}
 \label{width}
\ee
is its spreading along time, with $\tilde{\sigma}_t^0 = \sigma_0
\left[1 + (i\hbar t/2m\sigma_0^2)\right]$. However, due to the
friction, the wave packet  undergoes asymptotically (i.e., for $t
\to \infty$) a damping both in its propagation, according to
(\ref{free1}), stopping at the position $x_\infty = x_0 +
(p_0/m\gamma)$, and in its spreading, described by
\be
 \sigma_t = |\tilde{\sigma}_t| =
 \sigma_0 \sqrt{ 1 + \left( \frac{\hbar}{2m\sigma_0^2}\right)^2
   \left(\frac{1-e^{-\gamma t}}{\gamma}\right)^2} .
\ee
From this expression, we notice that the limit spreading is given
by $\sigma_\infty = \sigma_0 \sqrt{1 + (\hbar/2m\gamma\sigma_0^2)^2}$.

\begin{figure}[t]
 \begin{center}
  \includegraphics[width=16.5cm]{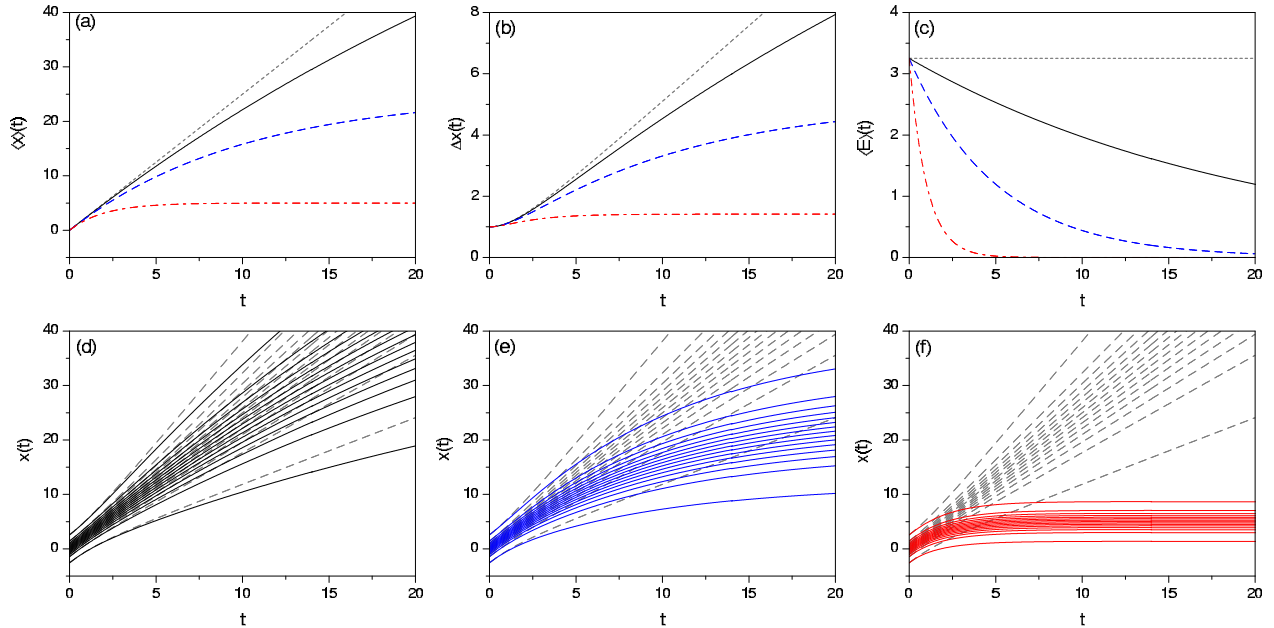}
  \caption{\label{fig1}
   Top: Average position (a), dispersion (b), and energy (c) for a
   Gaussian wave packet in free space affected by friction.
   Different values for the friction coefficient have been considered:
   $\gamma = 0.025$ (black solid line), $\gamma = 0.1$ (blue dashed
   line), and $\gamma = 0.5$ (red dash-dotted line).
   To compare with, the frictionless case ($\gamma = 0$) has also been
   included and is denoted with the gray dotted line.
   The value of the other parameters considered in these simulations
   were: $x_0=0$, $p_0=2.5$ ($E_0=3.25$), $\sigma_0=1$, $m=1$, and
   $\hbar=1$.
   Bottom: Bohmian trajectories associated with the three dissipative
   cases considered atop: (d) $\gamma = 0.025$, (e) $\gamma = 0.1$,
   and (f) $\gamma = 0.5$.
   Again, to compare with, the trajectories for the frictionless case
   (with the same initial conditions) have also been included in each
   panel (gray dashed lines).
   The initial positions have been distributed according to the initial
   Gaussian probability density.}
 \end{center}
\end{figure}

This dynamics is readily explained by inspecting the time-evolution
displayed by the corresponding Bohmian trajectories, which are obtained
after integration of the equation of motion (\ref{bohmdH}),
\be
  x(t) = x_t + \frac{\sigma_t}{\sigma_0}\ [x(0) - x_0] ,
 \label{trajdissipative}
\ee
which is formally equivalent to the expression that one obtains
for the free, frictionless case \cite{sanz:cpl:2007}. In the
latter case, illustrated in Fig.~\ref{fig1} for $\sigma_0 = 1$, $x_0=0$
and $p_0 = 2.5$ (arbitrary units), the initial boosting phase for short
times is followed by a linear evolution with time.
In the case with friction, we find that asymptotically,
Eq.~(\ref{trajdissipative}) reaches the limit
\be
 x(t \to \infty) = x_0 + \frac{p_0}{m\gamma}
  + \sqrt{1 + \left( \frac{\hbar}{2m\gamma\sigma_0^2} \right)^2}\
    [x(0) - x_0] .
 \label{frozenpos}
\ee
That is, the wave packet becomes localized: motionless and with
the spreading being frozen. Actually, in the case of strong
friction, 
it becomes essentially parallel to the classical or centroidal
trajectory, since the time-dependence vanishes very quickly and,
therefore, $\sigma_t$ becomes a constant value very rapidly. To
some extent, this is a step towards the classicality of the
quantum system without appealing to the more standard limiting
procedure of $\hbar\to 0$.

In Fig.~\ref{fig1} a series of wave-packet properties and Bohmian
trajectories are shown for different values of the friction
coefficient: $\gamma = 0.025$ (black solid line), $\gamma = 0.1$
(blue dashed line), and $\gamma = 0.5$ (red dash-dotted line).
These values correspond to an effective decrease of a 63.2\% of
the function at $t=40$, $t=10$, and $t=2$, respectively. In the
upper row, we show the average position (a), (spatial) dispersion
(b), and energy (c) as a function of time. To compare with, we
have also included the frictionless values, denoted with the gray
dotted line. The three quantities can be easily checked
analytically according to the relations found in the previous
section. Furthermore, the energy expectation value is given by
\be
 \bar{E} = \left( \frac{p_0^2}{2m}
  + \frac{\hbar^2}{8m\sigma_0^2} \right) e^{-2\gamma t} ,
\ee
i.e., it is suppressed at twice the rate $\gamma$. Bohmian
trajectories illustrating these dissipative cases are displayed in
the three lower panels, from left to right: (d) $\gamma = 0.025$,
(e) $\gamma = 0.1$, and (f) $\gamma = 0.5$. Again, to compare
with, the trajectories for the frictionless case have also been
included in each panel (gray dashed lines). To better appreciate
the different effect of the wave-packet spreading and how it is
suppressed as $\gamma$ increases, we have chosen 15~trajectories
with initial positions distributed according the initial Gaussian
probability density. Thus, for small frictions, we observe that
trajectories started in the ``wings'' of the wave packet will
increase faster their distance with respect to the centroid than
those closer to the latter. Moreover, this distance will be faster
for trajectories started behind the centroid than in front of it
due to the larger relative difference between their associated
velocities. This effect is, however, damped as the friction
coefficient increases, thus producing a smaller separation among
trajectories and eventually a freezing of their position for times
larger than $\gamma^{-1}$. These frozen positions are accounted
for by Eq.~(\ref{frozenpos}).

\subsection{Motion under a linear potential}
\label{sec4-2}

In the case of a linear potential, e.g., a gravitational or an electric
field, the classical solutions are also readily available.
Thus, if $V(x) = -max$ (with $a>0$, without loss of generality), we
find
\ba
 x_t & = & x_0
  + \frac{p_0}{m}\left(\frac{1 - e^{-\gamma t}}{\gamma}\right)
  + a\left(\frac{\gamma t - 1 + e^{-\gamma t}}{\gamma^2}\right) ,
 \label{ramp1} \\
 p_t & = & p_0 e^{-\gamma t}
   + ma\left(\frac{1 - e^{-\gamma t}}{\gamma}\right) ,
 \label{ramp2}
\ea
while $\alpha_t$ and $\gamma_t$ keep the same functional form as in the
free damped case (although $\mathcal{L}_t$ varies for the latter due to
the presence of a nonzero potential function), because only
second-order derivatives influence the evolution of these parameters
(through $V''_t$, as seen in (\ref{wave3})).

Again in this case, the frictionless limit $\gamma \to 0$ leads us
to the well-known expressions for uniform accelerated motion, with
$x_t = x_0 + (p_0/m)t + (a/2)t^2$ and $p_t = p_0 + mat$. The wave
packet also approaches the expression corresponding to the
solution to ramp-like potentials \cite{sanz:JPA:2011}, equal to
(\ref{wfsanz08}), except for the different functional form of
$E_t$ ($E_t = p_0^2/2m - max_0$) and an extra term coming from the
classical action. Regarding the long-time limit for finite
friction (i.e., for $t \gg \gamma^{-1}$), we find that while the
wave packet freezes its spreading, as in the previous example, it
still keeps moving due to the constant limit momentum, $p_\infty =
ma/\gamma$. Accordingly, the centroid of the wave-packet displays
a uniform motion described by
\be
 x_{t\to\infty} = x_0 + \frac{p_0}{m\gamma} - \frac{a}{\gamma^2}
  + \frac{a}{\gamma}\ t .
\ee

\begin{figure}[t]
 \begin{center}
  \includegraphics[width=16.5cm]{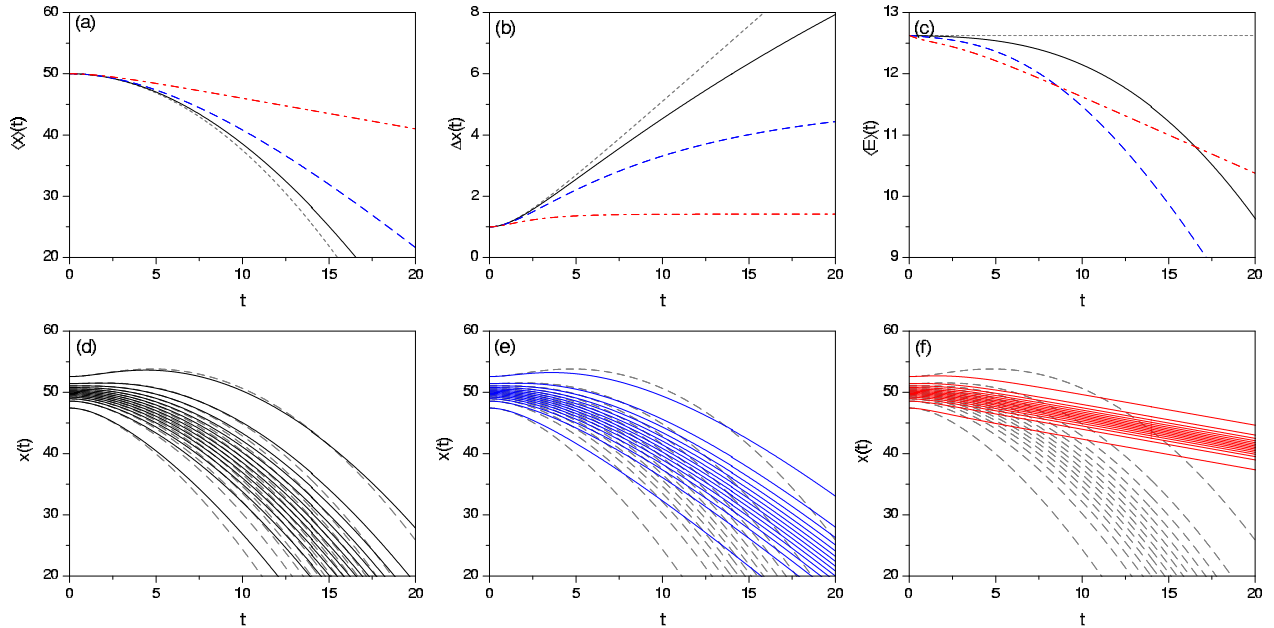}
  \caption{\label{fig2}
   Top: Average position (a), dispersion (b), and energy (c) for a
   Gaussian wave packet subject to the action of a linear potential,
   with $a=0.25$, and affected by friction.
   Different values for the friction coefficient have been considered:
   $\gamma = 0.025$ (black solid line), $\gamma = 0.1$ (blue dashed
   line), and $\gamma = 0.5$ (red dash-dotted line).
   To compare with, the frictionless case ($\gamma = 0$) has also been
   included and is denoted with the gray dotted line.
   The value of the other parameters considered in these simulations
   were: $x_0=50$, $p_0=0$ ($E_0=12.625$), $\sigma_0=1$, $m=1$, and
   $\hbar=1$.
   Bottom: Bohmian trajectories associated with the three dissipative
   cases considered atop: (d) $\gamma = 0.025$, (e) $\gamma = 0.1$,
   and (f) $\gamma = 0.5$.
   Again, to compare with, the trajectories for the frictionless case
   have also been included in each panel (gray dashed lines).
   The initial positions have been distributed according to the initial
   Gaussian probability density.}
 \end{center}
\end{figure}

These aspects are illustrated in the top panels of
Fig.~\ref{fig2}, where the average position (a), dispersion (b),
and energy (c) are plotted for the same values considered in the
previous section, and considering $a=0.25$. Moreover, the
frictionless homologous curves have also been included for
comparison (gray dotted line). As it can be noticed, as $\gamma$
increases, the transition from a uniformly accelerated motion,
displayed by the wave-packet centroid (see Fig.~\ref{fig2}(a)), to
a uniform rectilinear one, as a consequence of the damping,
becomes more apparent. The evolution towards this type of limit
motion, however, does not affect the width of the wave packet
differently from the case analyzed in the previous section, but it
appears very clearly in the average energy, particularly at large
values of $\gamma$. As it can be readily shown, the expression for
the average energy is given by
\be
 \bar{E} = \frac{p_0^2}{2m}\ e^{-2\gamma t} - max_0
  - p_0 a \gamma \left(\frac{1 - e^{-\gamma t}}{\gamma}\right)^2
  + \frac{ma^2}{2} \left(\frac{3 - 2\gamma t - 4e^{-\gamma t}
    + e^{-2\gamma t}}{\gamma^2}\right)
  + \frac{\hbar^2}{8m\sigma_0^2}\ e^{-2\gamma t} ,
 \label{eq6}
\ee
In the limit $\gamma t \gg 1$, this expression approaches
\be
 \bar{E} = - max_0 - \frac{p_0 a}{\gamma}
  + \frac{ma^2}{2\gamma^2} \left(3 - 2\gamma t\right) .
 \label{eq6b}
\ee
The linear regime typical of the limit motion can be inferred for
the lower values of $\gamma$. However, this effect is more
remarkable for larger $\gamma$, since the transient exponentials
vanish quickly and the energy immediately manifests its linear
dependence on time. Unlike the example of the free wave packet,
here the energy does not approach zero asymptotically, but it
continues decreasing below it as the wave packet slides downhill,
unless some additional constraint is imposed in the dynamics to
avoiding this situation. This is precisely the case of confining
potential functions displaying local minima, such as the harmonic
oscillator, which will be studied in the next section (for similar
behaviors in non-harmonic systems, for example, see
\cite{garashchuk:JCP:2013}). Nevertheless, although there is no
suppression of the motion, we can still observe spatial
localization.

To some extent the behavior displayed by the wave packet under the
action of a linear potential is similar, asymptotically, to reach
a classical-like regime. This can be better understood by
inspecting the associated Bohmian trajectories. These trajectories
are plotted in the lower panels of Fig.~\ref{fig2}, from (d) to
(e) for increasing values of $\gamma$. It is interesting that,
because $\alpha_t$ does not depend on the potential function, the
functional form displayed by the trajectories is exactly the same
as in the free case, except for the fact that these trajectories
contain information about the acceleration undergone by the
centroid (this information is transmitted through $x_t$). Now,
once the transients have vanished, all these trajectories evolve
parallel one another, just like classical trajectories under
similar circumstances (i.e., in this limit case).


\subsection{Motion in a damped harmonic oscillator}
\label{sec4-3}

Let us consider now the case of a harmonic oscillator potential,
$V(x) = m\omega_0^2 x^2/2$. In a frictionless situation, this
system is characterized by a series of eigenstates
\be
 \Phi_n(x,t) = N_n e^{-(m\omega_0/2\hbar)x^2 - i(n+1/2)\omega_0 t}
  H_n(\sqrt{m\omega_0/\hbar} x) ,
 \label{eigenHO}
\ee
with eigenenergy $E_n = (n + 1/2)\hbar\omega_0$, and where $H_n$
is the Hermite polynomial of degree $n$ and $N_n = (1/\sqrt{2^n
n!})(\pi\hbar/m\omega_0)^{-1/4}$ is the normalization constant.
These eigenvalues can be obtained by employing the same method
used above to derive the analytical expression of the time-evolved
wave packets \cite{heller:JCP:1975}. So, similarly, it can be used
to determine the dissipative counterpart of (\ref{eigenHO}),
\ba
 \Phi_n(x,t) & = &
  N_n e^{-(m\Omega/2\hbar)(1 + i\gamma/2\Omega)e^{\gamma t}x^2
  - i(n+1/2)\Omega t + \gamma t/4}
  H_n(\sqrt{m\Omega/\hbar}e^{\gamma t/2} x)
 \nonumber \\
 & = & N_n e^{-(m\Omega/2\hbar)e^{\gamma t}x^2 - i(n+1/2)\Omega t
   - i(m\gamma/4\hbar)e^{\gamma t}x^2 + \gamma t/4}
    H_n(\sqrt{m\Omega/\hbar}e^{\gamma t/2} x) .
 \label{eigenHOd}
\ea
However, contrary to (\ref{eigenHO}), Eq.~(\ref{eigenHOd})
describes a {\it quasi-stationary state}, i.e., states that, at
each time, are eigenstates of the dissipative Schr\"odinger
equation, but that eventually collapse to zero, as formerly shown
by Vandyck \cite{vandyck:JPA:1994}. Of course, these states are
only defined for $\omega_0 > \gamma/2$, as it will be explained
later on.

Concerning the associated Bohmian trajectories, from
(\ref{eigenHOd}), we obtain
\be
 \dot{x} = -\frac{\gamma}{2}\ x ,
 \label{eqmoteigen}
\ee
which after integration renders
\be
 x(t) = x(0) e^{-\gamma t/2} .
 \label{traeigen}
\ee
That is, regardless of the eigenstate considered, any trajectory
falls down to the bottom of the potential at the same rate,
$\gamma/2$, and therefore merging asymptotically at $x=0$. The
reason for such a behavior is that the model is fully dissipative
and there is no possibility for a feedback with an environment, as
happens when a Brownian-like motion is assumed. In this latter
case, the stochastic fluctuations accounting for the feedback with
a surrounding medium would be enough to sustain a dynamical regime
(even if stationary) and avoid its full collapse.

The previous case was already investigated by Vandyck
\cite{vandyck:JPA:1994} and in order to gain insight on it, now we
are going to consider the case of a coherent (or minimum
uncertainty) Gaussian wave packet initially centered around the
turning point $x=x_0$ ($p_0 = 0$). At any subsequent time, this
wave packet is described by
\begin{equation}
 \Psi(x,t) = \left( \frac{1}{2\pi\sigma_0^2} \right)^{1/4}
  e^{-(x-x_t)^2/4\sigma_0^2 + ip_t (x - x_t)/\hbar - i\omega_0 t/2
  + ip_t x_t/\hbar} ,
 \label{wpHO}
\end{equation}
with $x_t = x_0\cos \omega_0 t$ and $p_t = -m\omega_0 x_0 \sin
\omega_0 t$. The corresponding probability density reads as
\be
 |\Psi(x,t)|^2 = \frac{1}{\sqrt{2\pi\sigma_0^2}}\
  e^{-[x - x_0\cos(\omega_0 t)]^2/2\sigma_0^2} ,
\ee
where $\sigma_0^2 = \hbar/(2m\omega_0)$ for the wave packet (\ref{wpHO})
to be {\it coherent} (otherwise the wave packet will keep its Gaussian
shape, but it will display an oscillating variation of its width or
``breathing'' as it moves back and forth between the two turning points).
The corresponding quantum action is
\be
 S(x,t) = -\frac{1}{2}\ \hbar \omega_0 t
  - \frac{m\omega_0}{4}
  \left[4x x_0\sin(\omega_0 t) - x_0^2 \sin(2\omega_0 t)\right] ,
\ee
which leads to Bohmian trajectories oscillating with the same frequency
as $x_t$, but around their initial position, i.e.,
\be
 x(t) = [x(0) - x_0] + x_0 \cos(\omega_0 t) .
\ee

As also happens in classical mechanics, in order to proceed
analytically in the dissipative case, it is important to
distinguish three cases depending on whether $\omega_0$ is larger
than, equal to or smaller than $\gamma/2$, i.e., if we have
underdamped oscillatory motion, critically damped motion, or
overdamped motion, respectively. These situations lead to the
following solutions for the centroidal trajectory:
\ba
 \omega_0 > \gamma/2 & \Longrightarrow &
  \left\{ \begin{array}{rcl}
   x_t & = & \displaystyle
   \left(\frac{\omega_0}{\Omega}\right) x_0
    e^{-\gamma t/2} \cos (\Omega t - \varphi)
   \\
   p_t & = & \displaystyle
   - m \left(\frac{\omega_0^2}{\Omega}\right) x_0
    e^{-\gamma t/2} \sin \Omega t
  \end{array} \right. , \\
 \omega_0 = \gamma/2 & \Longrightarrow &
  \left\{ \begin{array}{rcl}
   x_t & = & \displaystyle
    x_0 \left( 1 + \frac{\gamma}{2}\ t \right) e^{-\gamma t/2}
   \\
   p_t & = & \displaystyle
    - m x_0 \left(\frac{\gamma}{2}\right)^2 e^{-\gamma t/2}
  \end{array} \right. , \\
 \omega_0 < \gamma/2 & \Longrightarrow &
  \left\{ \begin{array}{rcl}
   x_t & = & \displaystyle
   \left(\frac{\omega_0}{\Gamma}\right) x_0
    e^{-\gamma t/2} \cosh (\Gamma t + \phi)
   \\
   p_t & = & \displaystyle
   - m \left(\frac{\omega_0^2}{\Gamma}\right) x_0
    e^{-\gamma t/2} \sinh \Gamma t
  \end{array} \right. ,
\ea
where $\Omega = \sqrt{\omega_0^2 - (\gamma/2)^2}$, $\varphi =
(\tan)^{-1}(\gamma/2\Omega)$, $\Gamma = i\Omega$, and $\phi =
(\tanh)^{-1}(\gamma/2\Gamma)$. The relationship between $\omega_0$
and $\gamma$ also influences the calculation of $\alpha_t$ and
$f_t$. In the case of $\alpha_t$, the equation of motion to be
solved is
\be
 \dot{\alpha}_t = -\frac{2\alpha_t^2}{m}\ e^{-\gamma t}
   - \frac{1}{2}\ m\omega_0^2 e^{\gamma t} ,
\ee
which can be conveniently rearranged by introducing the change
$\alpha_t = g_t e^{\gamma t}$. This leads to the equation of
motion
\be
 \dot{g}_t = - \frac{2}{m}\ \left[ g_t^2 + \frac{m\gamma}{2}\ g_t
   + \left( \frac{m\omega_0}{2} \right)^2 \right]
   = -\frac{2}{m}\ (g_t - g_+) (g_t - g_-) ,
 \label{integd}
\ee
which does not contain exponential terms, and where
\be
 g_\pm = \frac{m}{2} \left[ - \frac{\gamma}{2}
    \pm \sqrt{\left(\frac{\gamma}{2}\right)^2 - \omega_0^2}\right] .
 \label{g1}
\ee
From the latter expression, it is now clear how the three cases of
damped motion also rule the behavior of $\alpha_t$, although there are
some physical subtleties to take into account.
If $\omega_0 \ne \gamma/2$, the general solution for $g_t$ reads as
\be
 g_t = \frac{g_+ (g_0 - g_-) e^{\beta t}
   - g_- (g_0 - g_+) e^{-\beta t}}
  {(g_0 - g_-) e^{\beta t} - (g_0 - g_+) e^{-\beta t}} ,
 \label{gmot}
\ee
where $\beta = \sqrt{(\gamma/2)^2 - \omega_0^2}$ and $g_0$ is the
initial condition. Otherwise, if $\omega_0 = \gamma/2$ (critically
damped motion), we have $g_+ = g_-$ and Eq.~(\ref{integd}) becomes
\be
 \dot{g}_t = -\frac{2}{m}\ (g_t - g_s)^2 ,
 \label{integd2}
\ee
with $g_s = -m\gamma/4$. The integration in time of this equation
of motion yields
\be
 g_t = g_s + \frac{g_0 - g_s}{1 + (g_0 - g_s)(2t/m)} .
 \label{gmot2}
\ee
If we now assume that our initial wave packet is (\ref{wpHO}) and
consider the initial condition $g_0 = \alpha_0 = \alpha_t =
im\omega_0/2$, then $g_t$ becomes an oscillatory function of time
for $\omega_0 > \gamma/2$, since $\beta = i\Omega$. Otherwise, it
becomes a monotonically decreasing function of time, with the
asymptotic limits $g_\infty = g_s$ if $\omega_0 = \gamma/2$, or
$g_\infty = g_+$ (with $\beta = \Gamma$) if $\omega_0 < \gamma/2$.

\begin{figure}[t]
 \begin{center}
  \includegraphics[width=16.5cm]{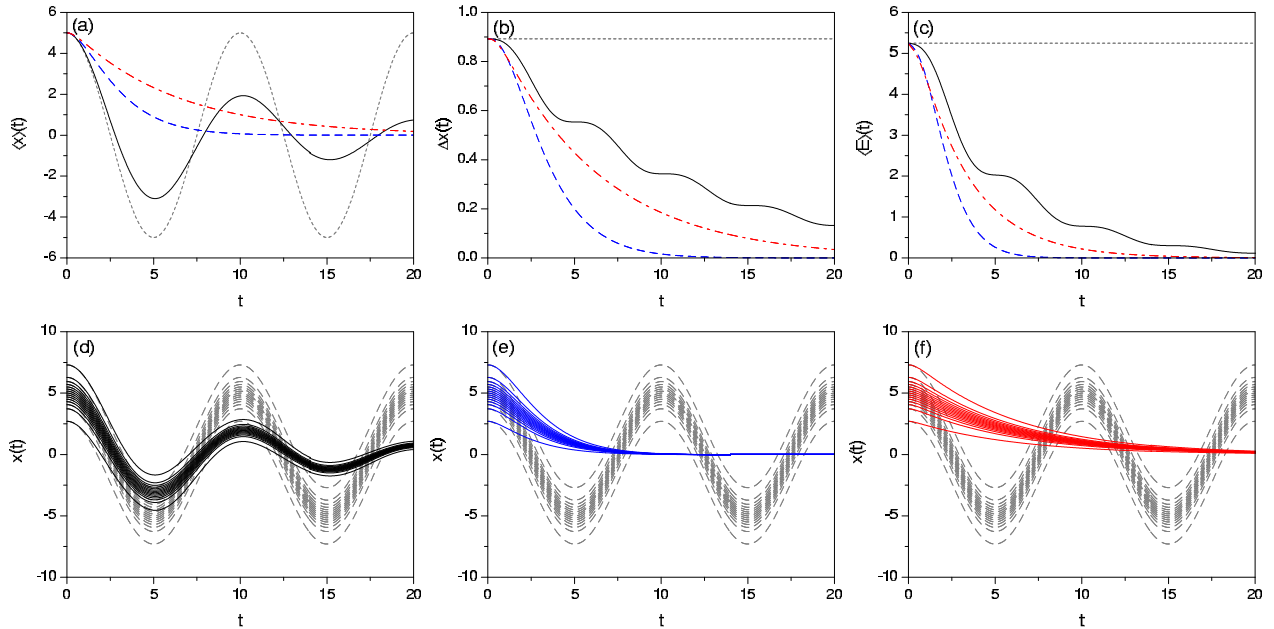}
  \caption{\label{fig3}
   Top: Average position (a), dispersion (b), and energy (c)
   for a coherent wave packet in a harmonic oscillator, with
   $\omega_0 = 2\pi/\tau_0 \approx 0.628$ ($\tau_0 = 10$), affected
   by friction.
   Values for the friction coefficient illustrative of the three
   regimes have been considered: $\gamma = 0.3\omega_0$ (black solid
   line), $\gamma = 2\omega_0$ (blue dashed line), and
   $\gamma = 4\omega_0$ (red dash-dotted line).
   To compare with, the frictionless case ($\gamma = 0$) has also been
   included and is denoted with the gray dotted line.
   The values of the other parameters considered in these simulations
   were: $x_0=5$, $p_0=0$ ($E_0 \approx 5.249$),
   $\sigma_0=\sqrt{\hbar/2m\omega_0}\approx 0.892$, $m=1$, and
   $\hbar=1$.
   Bottom: Bohmian trajectories associated with the three dissipative
   cases considered atop: (d) $\gamma = 0.3\omega_0$,
   (e) $\gamma = 2\omega_0$, and (f) $\gamma = 4\omega_0$.
   Again, to compare with, the trajectories for the frictionless case
   have also been included in each panel (gray dashed lines).
   The initial positions have been distributed according to the initial
   Gaussian probability density.}
 \end{center}
\end{figure}

In the top panels of Fig.~\ref{fig3}, we show the average position
(a), dispersion (b), and energy (c) for a wave function with the
initial form of the coherent state (\ref{wpHO}) in the three
dissipative regimes, with: $\gamma = 0.3\omega_0$ (black solid
line), $\gamma = 2\omega_0$ (blue dashed line), and $\gamma =
4\omega_0$ (red dash-dotted line). To compare with, the
corresponding frictionless values have also been included in each
panel (gray dotted lines). In the lower panels, we have included
the corresponding Bohmian trajectories, which effectively follow
the dissipative flow, falling down to the bottom of the well and
merging into the centroidal one.

We are not going to enter into more details here on the particular
analytical expressions for the quantities displayed in
Fig.~\ref{fig3}, as well as those for the corresponding Bohmian
trajectories. Nonetheless, they bring in an interesting question.
As seen in Fig.~\ref{fig3}(b), the decay of the dispersion or
width of the wave packet under underdamped conditions does not
follow a monotonic exponential-like decay, but displays an
oscillating behavior. Moreover, in the frictionless case, we saw
that the initial condition $\alpha_0 = im\omega_0/2$ led to a
time-independent width of the wave packet. An interesting analog
also arises in the dissipative case if we change the initial
condition. Notice in Eq.~(\ref{gmot}) that $g_t$ remains constant
with time if we assume that $g_0$ is equal to either $g_+$ or
$g_-$, which can be inferred either from (\ref{gmot}) or also
directly from (\ref{integd}) by setting $\dot{g}_t=0$. This means
$\alpha_t = g_+ e^{\gamma t}$, if $g_0 = g_+$, or $\alpha_t = g_-
e^{\gamma t}$, if $g_0 = g_-$. In order to choose the appropriate
solution, we consider the fact that in the limit $\gamma \to 0$
the chosen solution has to approach the non-dissipative value,
i.e., $\alpha_t \to im\omega_0/2$, which only happens if $g_0 =
g_+$. Therefore, we have that
\be
 \alpha_t = \frac{im\Omega}{2}
  \left( 1 + \frac{i\gamma}{2\Omega} \right) e^{\gamma t} .
 \label{alphat1}
\ee
Notice that in the limit of vanishing friction, this value
approaches the frictionless one mentioned above. More importantly,
$\alpha_t$ has a real and an imaginary part, and therefore the
wave function will be properly normalized. This can be readily
seen by integrating the equation of motion for $f_t$, which leads
to
\be
 f_t = \frac{i\hbar}{4}\
   \ln \left( \frac{\pi\hbar}{m\Omega} \right)
   - \frac{\hbar\Omega}{2} \left( 1 + \frac{i\gamma}{2\Omega} \right)t
   + \mathcal{S}_{{\rm cl},t} .
\ee
Substituting all these parameters in the functional expression for the
wave function, it reads as
\ba
 \Psi(x,t) & = &
 \left( \frac{\pi\hbar}{2{\rm Im}\{\alpha_0\}} \right)^{1/4}
  e^{-(m\Omega/2\hbar) (1 + i\gamma/2\Omega) e^{\gamma t}
    (x-x_t)^2 + ip_t (x - x_t)/\hbar
  - i\Omega t (1 + i\gamma/2\Omega)/2
  + i\mathcal{S}_{{\rm cl},t}/\hbar} ,
 \nonumber \\
  & = & \left( \frac{1}{2\pi\sigma_t^2} \right)^{1/4}
  e^{- (x-x_t)^2/4\sigma_t^2
  + ip_t (x - x_t)/\hbar - i\Omega t/2
    - (im\gamma/4\hbar) e^{\gamma t}(x-x_t)^2
    + i\mathcal{S}_{{\rm cl},t}/\hbar} ,
 \label{wpHOd}
\ea
with the width being $\sigma_t = e^{-\gamma
t/2}\sqrt{\hbar/2m\Omega}$. As it can be noticed, the first three
arguments in the exponential of (\ref{wpHOd}) are identical to
those in (\ref{wpHO}), but replacing $\omega_0$ by $\Omega$. The
wave function is properly normalized, but its width decreases
exponentially with time, according to
\be
 \Delta x = \sigma_t e^{-2\gamma t} .
\ee
Thus, it eventually collapses over the center of the harmonic well
as the whole wave packet follows the motion of a damped harmonic
oscillator. Along this transit, the system energy is also
exponentially lost, according to
\be
 \bar{E} = \frac{1}{2}\ m\omega_0^2 x_0^2
   \left( \frac{\omega_0}{\Omega} \right)^2
   \left[ 1 - \frac{\gamma}{2\omega_0}\
     \sin (2\Omega t - \varphi) \right] e^{-\gamma t}
  + \frac{1}{2}\ \omega_0 \hbar \left(\frac{\omega_0}{\Omega}\right)
    e^{-\gamma t} .
\ee

It is worth stressing that for underdamped conditions,
$\alpha_t$, given by (\ref{alphat1}), is a complex function, which
makes the wave function (\ref{wpHOd}) to display a vanishing
Gaussian shape plus a phase factor as the system oscillates, as
seen in (\ref{wpHOd}). Critically damped and overdamped conditions
imply that the motion amplitude of the system exhibits a monotonic
decrease to zero, with no oscillations. In the present context,
where we are seeking for solutions such that $\dot{g}_t=0$, this
now translates into a rather puzzling quantum behavior. In order
to analyze it, let us start from the overdamped condition,
$\omega_0 < \gamma/2$. Equation~(\ref{gmot}) is still valid,
although we have $g_\pm = \pm (m\Gamma/2)(1 \mp \gamma/2\Gamma)$.
Between these two solutions, we choose $g_+$, because it is
consistent with (\ref{alphat1}) and also because it corresponds to
the long-time limit of $g_t$ in this case (i.e., for $\beta =
\Gamma$). Accordingly, the spreading factor will read as
\be
 \alpha_t = \frac{m\Gamma}{2} \left(1 - \frac{\gamma}{2\Gamma}\right)
  e^{\gamma t} ,
\ee
which is a pure real function.
From it,
\be
 f_t = f_0
   + \frac{i\hbar\Gamma}{2} \left( 1 - \frac{\gamma}{2\Gamma} \right)t
   + \mathcal{S}_{{\rm cl},t} ,
\ee
where $f_0$ is now left as a free parameter, although its
imaginary part has to be such that the initial plane wave is still
normalized and validates the working hypothesis regarding the
expectation values of the position and momentum. Notice that in
this case, the corresponding wave function reads as
\be
 \Psi(x,t) =
  e^{-(im\Gamma/2\hbar)(1 + \gamma/2\Gamma) e^{\gamma t}(x-x_t)^2
   + ip_t (x - x_t)/\hbar + \Gamma t (1 + \gamma/2\Gamma)/2
   + if_0 + i\mathcal{S}_{{\rm cl},t}/\hbar} ,
 \label{wpHOd2}
\ee
which is a pure phase factor multiplied by a diverging
time-dependent exponential. It is interesting, though, that
although the wave function itself diverges, the associated Bohmian
trajectories are well-defined and approach asymptotically the
classical overdamped centroid, $x_t$ (see below).

Regarding the critically damped situation, if we consider here the
initial value $g_0 = g_s$, the stationary solution is $g_t = g_s =
-m\gamma/4$, and therefore
\be
 \alpha_t = - \frac{m\gamma}{4}\ e^{\gamma t} .
\ee
As it can be seen, this value of $\alpha_t$ allows us to connect
smoothly those two previously obtained in underdamped and overdamped
cases as the friction $\gamma$ is gradually increased.

In spite of the type of motion displayed in each one of the regimes
discussed above, the Bohmian equation of motion for the trajectories
can be recast in the same form for all of them,
\be
 \frac{d(x - x_t)}{x - x_t} = -\frac{\bar{\gamma}}{2}\ dt ,
\ee
where $\bar{\gamma} = \gamma$ for underdamped and critically
damped motions, and $\bar{\gamma} = \gamma - 2\Gamma$ for
overdamped motions. As it can be noticed, this equation looks
exactly the same as Eq.~(\ref{eqmoteigen}) for the eigenstates,
although replacing $x$ by $x-x_t$. Thus, the solutions,
\be
 x(t) = x_t + [x(0) - x_0] e^{-\bar{\gamma} t/2} ,
 \label{tracoh}
\ee
are also very similar. In the case of Eq.~(\ref{traeigen}), since
the associated wave function is a quasi-eigenstate, the trajectory
approaches asymptotically the origin or, in other words, it falls
down to the bottom of the well. In the case of Eq.~(\ref{tracoh}),
and differently to what we have seen in the two previous sections,
any trajectory will eventually coalesce with the centroidal one.
This is an interesting case where the Bohmian non-crossing rule
seems to be violated. However, we have to keep in mind that we are
describing an effective dissipative dynamics, which only models
phenomenologically the system behavior, but now its true
interaction with the environment causing such an effect. Actually,
the evolution of the trajectories is in compliance with the
shrinking undergone by the wave function in the damped oscillatory
regime and its evanescent nature in the overdamped regime. These
trajectories, described generically by Eq.~(\ref{tracoh}), have
the following explicit functional form
\be
 x(t) = \left\{ \begin{array}{lcc}
  \left\{ \displaystyle [x(0) - x_0]
  + x_0 \left( \frac{\omega_0}{\Omega} \right) \cos(\Omega t - \varphi)
  \right\} e^{-\gamma t/2} , & & \omega_0 > \gamma/2 \\
  \left\{ \displaystyle [x(0) - x_0]
  + x_0 \left( 1 + \frac{\gamma t}{2} \right)
  \right\} e^{-\gamma t/2} , & &  \omega_0 = \gamma/2 \\
  \left\{ \displaystyle [x(0) - x_0] e^{\Gamma t}
  + x_0 \left( \frac{\omega_0}{\Gamma} \right) \cosh(\Gamma t + \phi)
  \right\} e^{-(\gamma - 2\Gamma) t/2} , & & \omega_0 < \gamma/2
  \end{array} \right. ,
\ee
where the first expression looks very similar to that for the
unperturbed harmonic oscillator, with the exception of the overall
exponentially decaying factor and the dephasing in the cosine function.


\subsection{Interference dynamics}
\label{sec4-4}

Now, consider a problem involving wave-packet interference, for which
a general solution can be expressed as a coherent superposition
\cite{sanz:JPA:2008},
\be
 \Psi = \mathcal{N} (\Psi_1 + \Psi_2) ,
\ee
with $\mathcal{N}$ being the overall norm factor (it is assumed that
each Gaussian wave packet, given by (\ref{wfsanz08}), is normalized).
As it can be shown, the associated Bohmian equation of motion reads as
\ba
 \dot{x} & = &
  \frac{\rho_1}{\rho}\ \dot{x}_1 + \frac{\rho_2}{\rho}\ \dot{x}_2
  + \frac{\hbar}{2mi\rho} \left( \Psi_1^* \partial_x \Psi_2
  - \Psi_2 \partial_x \Psi_1^* \right) e^{-\gamma t}
  + \frac{\hbar e^{-\gamma t}}{2mi\rho}
  \left( \Psi_2^* \partial_x \Psi_1
  - \Psi_1 \partial_x \Psi_2^* \right) e^{-\gamma t}
 \nonumber \\
  & = &
  \frac{\rho_1}{\rho}
  \left[\frac{p_{t,1}}{m}
  + \frac{2{\rm Re}\{\alpha_{t,1}\}}{m}\ (x-x_{t,1})\
   e^{-\gamma t}\right]
  + \frac{\rho_2}{\rho}
  \left[\frac{p_{t,2}}{m}
  + \frac{2{\rm Re}\{\alpha_{t,2}\}}{m}\ (x-x_{t,2})\
   e^{-\gamma t}\right]
 \nonumber \\ & &
  + 2\cos \xi_{12}\ \frac{\sqrt{\rho_1\rho_2}}{\rho}
  \left[\frac{p_{t,1} + p_{t,2}}{2m}
  + \left[\frac{{\rm Re}\{\alpha_{t,1}\}(x-x_{t,1})
    + {\rm Re}\{\alpha_{t,2}\}(x-x_{t,2})}{m} \right]
    e^{-\gamma t} \right]
 \nonumber \\ & &
  - 2\sin \xi_{12}\ \frac{\sqrt{\rho_1\rho_2}}{\rho}
  \left[\frac{{\rm Im}\{\alpha_{t,1}\}(x-x_{t,1})
    - {\rm Im}\{\alpha_{t,2}\}(x-x_{t,2})}{m} \right]
    e^{-\gamma t} ,
 \label{traj2wp}
\ea
where $\dot{x}_i$ refers to the equation of motion for the $i$th
wave packet and $\xi_{12} = (S_1 - S_2)/\hbar$, being $\rho_i$ and
$S_i$ the probability density and real phase associated with the
$i$th wave packet when it is expressed in polar form. As it can be
noticed, this expression simply reflects the sum of two separate
fluxes, each one associated with one wave packet, plus another one
coming from their interference.

\begin{figure}[t]
 \begin{center}
  \includegraphics[width=16.5cm]{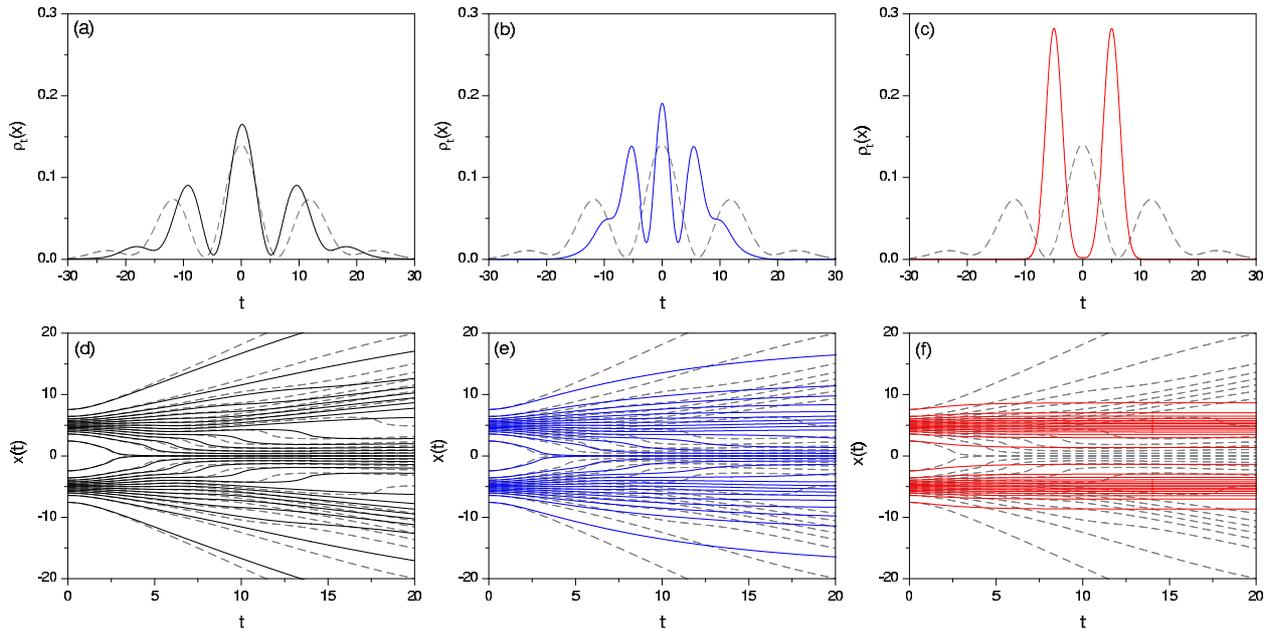}
  \caption{\label{fig4}
   Top: Final probability density (at $t=20$) for a coherent
   superposition of two Gaussian wave packets in free space
   affected by friction.
   The same friction values as in Fig.~\ref{fig1} have been considered:
   (a) $\gamma = 0.025$ (black solid line), (b) $\gamma = 0.1$ (blue
   solid line), and (c) $\gamma = 0.5$ (red solid line).
   To compare with, the frictionless case ($\gamma = 0$) has also been
   included in each panel (gray dashed line).
   The value of the other parameters considered in these simulations
   were: $x_{0,1}=5=-x_{0,2}$, $p_{0,1}=p_{0,2}=0$ ($E_0=3.25$),
   $\sigma_{0,1}=\sigma_{0,2}=1$, $m=1$, and $\hbar=1$.
   Bottom: Bohmian trajectories associated with the three dissipative
   cases considered atop: (d) $\gamma = 0.025$, (e) $\gamma = 0.1$,
   and (f) $\gamma = 0.5$.
   Again, to compare with, the trajectories for the frictionless case
   (with the same initial conditions) have also been included in each
   panel (gray dashed lines).
   The initial positions have been distributed according to the initial
   Gaussian probability density.}
 \end{center}
\end{figure}

Depending on the value given to the parameters of each wave
packet, one can generate different dynamics. For example, if they
have the same width ($\alpha_{t,1} = \alpha_{t,2} = \alpha_t$) and
are located at symmetric positions with respect to $x=0$ ($x_{t,1}
= x_0 = - x_{t,2}$), then (\ref{traj2wp}) simplifies to
\ba
 \dot{x} & = &
  \frac{\rho_1}{\rho}
  \left[\frac{p_{t,1}}{m}
  + \frac{2{\rm Re}\{\alpha_t\}}{m}\ (x-x_0) e^{-\gamma t} \right]
  + \frac{\rho_2}{\rho}
  \left[\frac{p_{t,2}}{m}
  + \frac{2{\rm Re}\{\alpha_t\}}{m}\ (x+x_0) e^{-\gamma t} \right]
 \nonumber \\ & &
  + 2\cos \xi_{12}\ \frac{\sqrt{\rho_1\rho_2}}{\rho}
  \left[\frac{p_{t,1} + p_{t,2}}{2m}
  + \frac{2{\rm Re}\{\alpha_t\}}{m}\ x e^{-\gamma t} \right]
  + 2\sin \xi_{12}\ \frac{\sqrt{\rho_1\rho_2}}{\rho}
  \left[\frac{2{\rm Im}\{\alpha_t\}}{m}\ x_0 e^{-\gamma t} \right] .
 \label{traj2wp2}
\ea
Thus, consider the case of two interfering coherent wave packets
in free space. This situation, for example, mimics the situation
of a two slit experiment when each wave packet is associated with
one of the diffracted beams. In such a case, if the longitudinal
propagation (parallel to the diffractive screen) is slower than
the perpendicular one (in the direction of the diffracted beam),
both motions can be decoupled and treated as two independent
one-dimensional propagations, one longitudinal and the other
translational, the latter being well accounted for by a plane
wave. Now, if one assumes that $p_{0,1} = p_{0,2} = 0$, i.e.,
there is no translational motion along the longitudinal direction,
but only spreading of the two wave packets \cite{sanz:JPA:2008},
the above expression (\ref{traj2wp2}) will read as
\ba
 \dot{x} & = &
  \frac{\rho_1}{\rho}
  \left[\frac{2{\rm Re}\{\alpha_t\}}{m}\ (x-x_0)\ e^{-\gamma t}\right]
  + \frac{\rho_2}{\rho}
  \left[\frac{2{\rm Re}\{\alpha_t\}}{m}\ (x+x_0)\ e^{-\gamma t}\right]
 \nonumber \\ & &
  + 2\cos \xi_{12}\ \frac{\sqrt{\rho_1\rho_2}}{\rho}
  \left[\frac{2{\rm Re}\{\alpha_t\}}{m}\ x e^{-\gamma t} \right]
  + 2\sin \xi_{12}\ \frac{\sqrt{\rho_1\rho_2}}{\rho}
  \left[\frac{2{\rm Im}\{\alpha_t\}}{m}\ x_0 e^{-\gamma t} \right] .
 \label{traj2wp2free}
\ea
Results after integration of this equation of motion for different
values of $\gamma$ (the same values as in the case of the free
wave packet treated above) can be seen in Fig.~\ref{fig4}. As it
can be seen through the sets of trajectories selected, the
friction of the medium leads to the localization of the two wave
packets by gradually ``freezing'' them rather than by annihilating
the coherence between them due to the interaction with an
environment, as happens with
\cite{sanz:EPJD:2007,sanz:CPL:2009-2}. In other words, the reason
why interference is not observed in a quantum viscid medium is
because the wave packets cannot be seen each other, rather than
because the destruction (over time) of their mutual coherence.

\begin{figure}[t]
 \begin{center}
  \includegraphics[width=16.5cm]{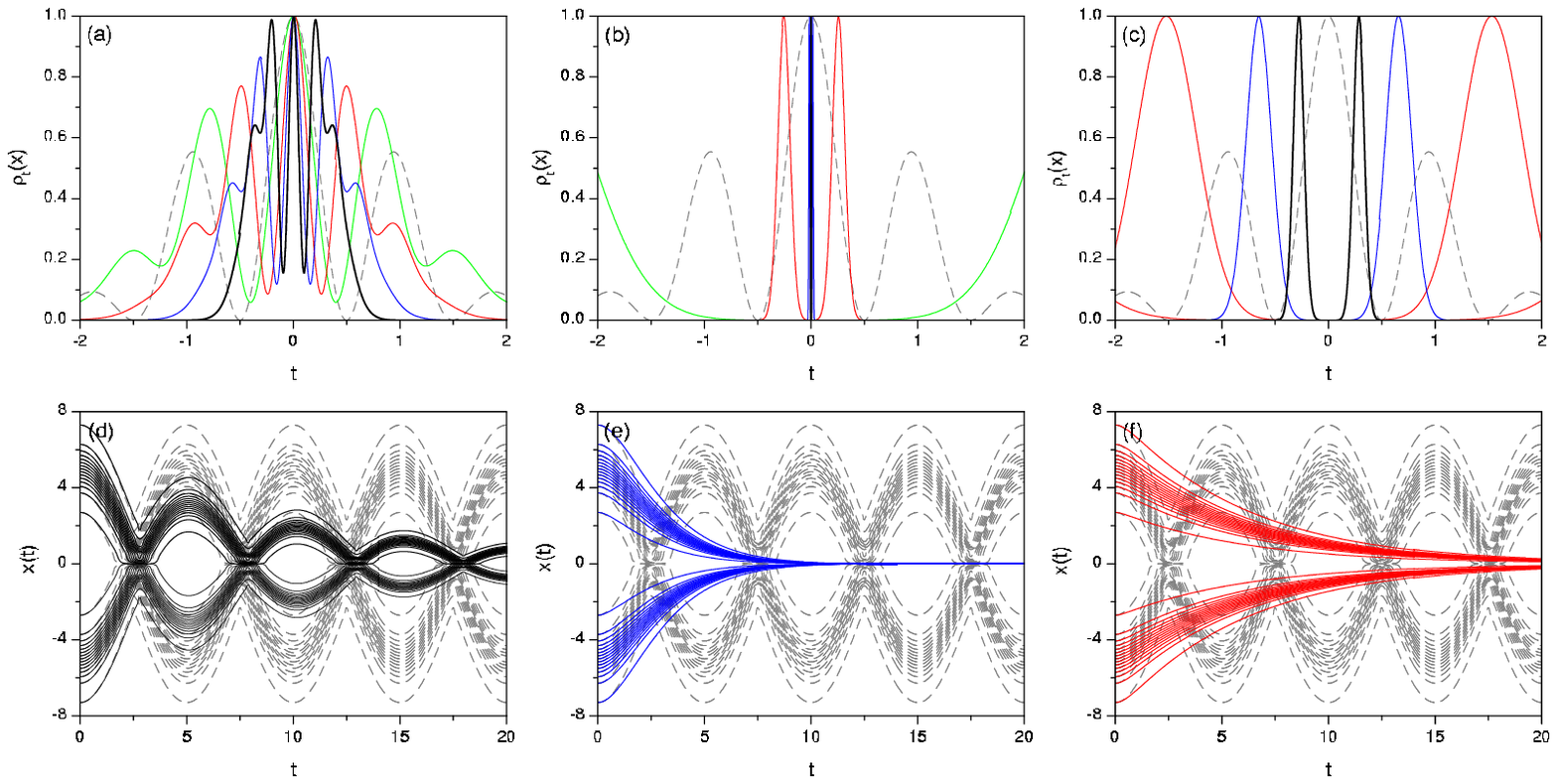}
  \caption{\label{fig5}
   Top: Probability density at different times for a coherent
   superposition of two Gaussian wave packets (coherent states) in a
   harmonic oscillator, with $\omega_0 = 2\pi/\tau_0 \approx 0.628$
   ($\tau_0 = 10$), affected by friction.
   The same friction values as in Fig.~\ref{fig3} have been considered:
   (a) $\gamma = 0.3\omega_0$, (b) $\gamma = 2\omega_0$, and (c)
   $\gamma = 4\omega_0$.
   In each panel, in order to better appreciate the damping effects,
   the probability density at times of maximal interference in the
   frictionless case is shown: $t=\tau_0/2$ (green solid line),
   $t=3\tau_0/2$ (red solid line), $t=5\tau_0/2$ (blue solid line),
   and $t=7\tau_0/2$ (black thick solid line); to compare with, the
   frictionless ($\gamma = 0$) probability density at these times has
   also been included in each panel (gray dashed line).
   Moreover, to better appreciate the effect of friction with time,
   the maximum for all probability densities has been re-scaled to 1.
   Bottom: Bohmian trajectories associated with the three dissipative
   cases considered atop: (d) $\gamma = 0.3\omega_0$,
   (e) $\gamma = 2\omega_0$, and (f) $\gamma = 4\omega_0$.
   Again, to compare with, the trajectories for the frictionless case
   (with the same initial conditions) have also been included in each
   panel (gray dashed lines).
   The initial positions have been distributed according to the initial
   Gaussian probability density.}
 \end{center}
\end{figure}

Another situation of interest is when the two wave packets are
inside a harmonic oscillator, each one launched from an opposite
turning point (i.e., again $x_{0,1} = x_0 = - x_{0,2}$). In this
case, $x_{t,1} = - x_{t,2}$ and $p_{t,1} = - p_{t,2}$, so the
general form of equation of motion (\ref{traj2wp2free}) remains
still valid, although also keeping the two momentum terms that
appear in the first line of (\ref{traj2wp2}). This dissipative
dynamics is illustrated in Fig.~\ref{fig5} for the same cases
treated above for the harmonic oscillator. As it can be seen,
because of the presence of the oscillator, both wave packets
approach the center of the well, but without ever crossing it: in
the underdamped case a series of bounces are observed until the
trajectories merge into the single asymptotic one at $x=0$, while
for the critically damped and overdamped cases the two sets of
trajectories gradually merge into this asymptotic one. Again here
we find an example of spatial localization, although it is
different from the free case seen above, since the final wave
function becomes a $\delta$-like function regardless of where the
trajectories came from, whether from one turning point or the
other.

\section{Concluding remarks}
\label{sec5}

The approach presented here to deal with dissipation in quantum viscid
media is based on Bohmian mechanics.
However, an even more direct link can
be established with the hydrodynamic reformulation of
Schr\"odinger's wave mechanics proposed in 1926 by Madelung
\cite{madelung:ZPhys:1926}. Accordingly, the evolution of a quantum
system can be connected to that of an ideal (quantum) fluid by means
of a simple nonlinear transformation, which goes from the wave
function to two real fields, namely the probability density and the
phase field (or, to make the analogy more apparent, the quantum
current density). This work would then go a step beyond by assuming
the medium is not ideal, but viscid. To some extent, the
introduction of a friction makes the approach to somehow resemble
Bell's beables interpretation, particularly as further developed by
Vink \cite{vink:PRA:1993} and Lorenzen {\it et al.}\
\cite{lorenzen:PRA:2009}. Actually, the latter also considered
friction as a means to interpret and understand phenomena such as
dissipation, decoherence, or the quantum-to-classical transition in
quantum systems.

Now, the problem posed by the Caldirola-Kanai model analyzed here is
that the standard quantum uncertainty relations are violated.
Alternative nonlinear models have been proposed in the literature to
overcome this inconvenience, such as Kostin's model \cite{kostin:jcp:1972},
which includes in the Schr\"odinger equation a friction term proportional
to the phase of the wave function (related, as seen above, to the
Bohmian momentum).
Because of this term, although the quantum system displays a similar
damping to the one observed here, the width of the wave function does
not collapse to zero in the case of the harmonic oscillator, for
example, thus preserving the uncertainty relations.
More importantly, this model has been recently used by Garashchuk and
coworkers \cite{garashchuk:JCP:2013} with the practical purpose of
obtaining the ground state in arbitrary potential functions ---an
analogous result, starting
from similar grounds, was previously found by Doebner and Goldin
\cite{doebner:pla:1992}.
The difference between the effective
dissipative Schr\"odinger equation found by the latter authors with
respect to Eq.~(\ref{dissschro}) arises from the way how it is obtained
from the Langevin equation (\ref{langed}).
While in our case this is done from a linear, but time-dependent
Hamiltonian operator, in the case of these authors the Hamiltonian
is time-independent, but nonlinear ---an interesting discussion on
these two procedures can be found in \cite{schuch:IJQC:1999}.
From a physical viewpoint, the role of this nonlinearity is equivalent
to adding a noise on the right-hand side of Eq.~(\ref{langed}) in order
to avoid the collapse of the system or its localization.
That is, it enables a kind of fluctuation-dissipation relation, although
the corresponding nonlinear Schr\"odinger equation (Kostin's equation)
still describes one subsystem and, therefore, the associated Bohmian
trajectories are reduced in the same sense as those here analyzed
(we still have an incomplete view of the process).

Alternative and possible applications of this type of studies can
be developed in several ways. First, it is well known that a
quantum Brownian particle moving in a periodic potential, coupled
to a dissipative environment and at zero temperature, undergoes a
transition from an extended to a localized ground state as the
friction is raised \cite{hakim:prl:1985,fisher:prb:1985}. We think
our methodology could throw some new light on this aspect as well
as when a harmonic oscillator is interacting with a
one-dimensional massless scalar field \cite{zurek:prd:1989}.
Second, when considering the lowest frequency motion or frustrated
translational motion of adsorbates on corrugated surfaces
\cite{sanz-bk-1}. In this problem, such a motion can be similarly
seen as that described by a damped harmonic oscillator. It is true
that a certain feedback between the thermal bath due to the
presence of the surface and the quasi harmonic oscillator cannot
be neglected. However, if the surface temperature is really low,
the corresponding dynamics has to be closer to a pure dissipative
dynamics as considered in this work. Third, within the same
context, the diffusion of adsorbates on flat surfaces implies a
free propagation of wave packets in the presence of friction. Again,
the surface temperature is supposed to be low enough in order to
highlight the role of the dissipation. In a flat surface, the
collision between two adsorbates could also be described by the
interference dynamics in presence of a thermal bath or surface. In
this sense, a new friction mechanism could be added which is the
so-called collisional friction \cite{sanz-bk-1}. And fourth, our
study could be easily extended to gas phase molecular processes
where the interaction among different molecules or atoms due to a
high pressure could be replaced by such a friction mechanism. It
should be then easily analyzed, for example,  the broadening of
spectral lines in this collisional regime.


\section*{Acknowledgements}

This work has been supported by the Ministerio de Econom{\'\i}a y
Competitividad (Spain) under Project FIS2011-29596-C02-01;
AS is also grateful for a ``Ram\'on y Cajal'' Research Grant.




\end{document}